\documentclass[letterpaper, 10 pt, conference]{ieeeconf}

\IEEEoverridecommandlockouts                              

\usepackage{graphics} 
\usepackage{epsfig} 
\usepackage{mathptmx} 
\usepackage{times} 
\usepackage{amsmath} 
\usepackage{amssymb}  
\usepackage{epstopdf}
\usepackage{color}
\usepackage{indentfirst}
\usepackage{subcaption}
\title{\LARGE \bf
	Power Regulation in High Performance Multicore Processors$^{\dag}$
\thanks{$^{\dag}$Research  supported in part by the NSF under Grant CNS-1239225.}
}

\author{X. Chen,  Y. Wardi, and S. Yalamanchili$^{*}$
	\thanks{$^*$School of Electrical and Computer Engineering, Georgia Institute of Technology, Atlanta, GA 30332. Email:
		xchen318@gatech.edu,  ywardi@ece.gatech.edu, sudha@ece.gatech.edu.}
}

\begin{document}
\maketitle
\thispagestyle{empty}
\pagestyle{empty}

\begin{abstract}
This paper presents, implements, and evaluates a power-regulation technique
for multicore processors, based on an integral controller with
adjustable gain. The gain is designed for wide stability margins,
and computed in real time as part of the control law. The
tracking performance of the control system is robust with
respect to  modeling uncertainties and computational errors in
the loop. The main challenge of designing such a controller is
that the power dissipation of program-workloads
varies widely and often cannot
be measured accurately; hence extant controllers are either ad hoc or
based on a-priori modeling characterizations of the processor and workloads.
Our approach is different. Leveraging the aforementioned
robustness it uses a simple textbook modeling framework, and
adjusts its parameters in real time by a system-identification
module. In this it trades modeling precision for fast computations
in the loop making it suitable for on-line implementation in commodity data-center processors.
Consequently, the proposed controller is
agnostic in the sense that it does not require any a-priori
system characterizations. We present an implementation of
the controller on Intel's fourth-generation microarchitecture,
Haswell, and test it on a number of industry benchmark programs
which are used in scientific computing and datacenter
applications. Results of these experiments are presented
in detail exposing the practical challenges of implementing
provably-convergent  power regulation solutions in commodity multicore processors.

\end{abstract}

%



\section{Introduction}

For decades, scaling of transistors to decreasing geometries was the
primary source of increased processor performance. This was
accompanied by the corresponding scaling of device power thereby
keeping power densities roughly constant on a processor die. However,
this behavior known as Dennard scaling has ended leading to
unsustainable growth in power consumption in future processors as we
increase the number of transistors on a die. Therefore, to continue
to sustain performance scaling we must seek new and innovative
advances in power management in multicore processors.  Such advances
are central to the effective operation of all modern processors in
platforms ranging from mobile devices to data centers and
high-performance computing (HPC) machines that drive national
initiatives in key areas such as science, finance, and
defense~\cite{Wang11,Bohrer02}.

In multicore processors, the relationships between workloads,
power dissipation, resulting thermal fields, and their interaction
with the leakage current present new and unresolved power and thermal
management challenges. For example, application workloads exhibit
time-varying computation and memory access behaviors resulting in
spatially and temporally varying power dissipation and non-uniform thermal
fields. The cross-chip variations in temperature couples to circuit
leakage and delay, increases full-chip leakage power, reduces peak
throughput, degrades chip/package reliability, and increases
cooling/packaging costs. Thus, effective control of power dissipation
is critical to the reliable and high performance operation of multicore
processors. This paper describes a novel and effective power
regulation technique and the results of an evaluation of its
implementation on a commodity multicore processor.

Modern multicore processors are organized into several {\em voltage
  islands} where each island may comprise of one or more processing
cores. Each voltage island can operate at one of several discrete {\em
  power states} each defined by an operational voltage-frequency
pair. A general technique for controlling power and temperature is
based on setting the appropriate power state of each voltage island.
This is commonly referred to as Dynamic Voltage/Frequency Scaling, or
DVFS. The development of effective controls based on DVFS faces
several challenges.  First, the relationship between the clock
frequency and core power is complicated by other factors such as the
coupling between temperature and leakage power.  Second, application
workloads have time-varying compute and memory system behaviors
requiring a robust, adaptive control strategy to manage power
dissipation.  Third, distinct cores in a voltage island execute
distinct instruction streams with distinct behaviors but may share a
common clock and hence frequency.  For example, the Intel Haswell
processor tested in this paper has four cores sharing a single voltage
island and  executing eight hardware threads (subprograms) at the same
frequency~\cite{Hammarlund14}.

A number of DVFS heuristics have been proposed to control power
dissipation. Prominent are heuristics for clock
gating~\cite{Shadron03}, thread migration~\cite{Liu12}\cite{Yeo08},
prediction \cite{Kim15} and voltage
scaling~\cite{Avirneni16}. However, heuristics are limited in their
scope and robustness. Consequently, several efforts have applied
feedback control theory as an effective way to improve performance and
robustness~\cite{Raghavendra08,Mishra10,Hellerstein04}. Generally,
such controllers relied on off-line analysis of anticipated
workloads~\cite{Mishra10,Wang11} or empirical
approaches~\cite{Deval15,Krishnaswamy15} to derive control
parameters. This includes efforts to limit operation below a maximum
power constraint~\cite{Chen13,Lefurgy08}. However, all of these
approaches are applied to applications that have been profiled a
priori to derive the control parameters.

This paper concerns a control law that is not based on any off-line
profiling (hence said to be {\it agnostic}), and it estimates the
model-parameters on line by  least-square system identification.
The control law is
comprised of a standalone integrator with adaptive gain. Now it is
well-known that an integral control can have poor stability margins
and oscillations in the system's response, hence it is often
supplemented by proportional and derivative elements in order to form
the PID control \cite{Franklin14}. We use a different approach, consisting of a
standalone integrator with a variable gain, designed for fast
convergence, wide stability margins, and reduced
oscillations as compared to fixed-gain integrators. Moreover, its tracking performance and stability are
quite robust to modeling variations and computing errors in the loop,
and hence we need not worry about precise model
parameters. Furthermore, we can speed up the computations in the loop
at the expense of precision if needed.\footnote{Refs. \cite{Almoosa12,Wardi16} argue for the choice of the variable-gain integral control described in this paper over a PID controller, based on its robustness and flexibility in implementation. Furthermore,
 we have tested via simulation (not reported here) the
 addition of a proportional element to the integral controler but  found no improvement.
}

The controller described in the sequel was first designed for
regulating the dynamic power in computer cores in Ref. \cite{Almoosa12}.
Subsequently it has been analyzed in an abstract setting in \cite{Wardi16},
where its convergence, stability, and robustness were proved.  Its
performance was tested via simulation on instruction-throughput
regulation \cite{Almoosa12a,Chen15}, then on throughput of abstract Discrete Event
Dynamic Systems (DEDS) such as queues, Petri nets  and
transportation networks \cite{Seatzu14}. Lately the controller has been
implemented on Intel's fourth-generation micro-architecture, Haswell
\cite{Chen16}, where it was tested on instruction-throughput regulation.

This paper concerns an implementation of the controller on a Haswell
machine and evaluations of its application to power regulation.
It makes the following specific contributions:
1). It is the
first to present an implementation of an integral control for power
regulation in multicore processors. 2).  It is
agnostic, and adjusts well to workload variations. 3).
 It is the first (to our knowledge)
to use on-line system identification for estimating a suitable system-model.  4). It is applied
to timely data-center applications.  5). It converges quite fast.

The rest of the paper is organized as follows. Section II   describes
the problem, system-model, and power-regulation technique, and recounts established results.
Section III presents test results of applications of the controller to industry-benchmarks, and
Section IV concludes
the paper.

\section {Power Regulation Technique}

This section first describes the regulation technique in  the abstract setting considered in
\cite{Wardi16} in order to  highlight its
general salient features. Then it discusses its particular applications to  power control in multicore
 processors.

Consider the single-input-single-output discrete-time system shown in Figure 1, where $k=1,2,\ldots$ represents discrete time, $r\in R$
is a reference input, $y_k\in R$ is the output, $u_k$ is the control variable, and $e_k\in R$ is the error signal.
Generally the plant can be nonlinear and time varying, and the objective of the controller is to have the output
$y_k$, $k=1,2,\ldots$, track the reference $r$.

\begin{figure}	[!t]
	\centering
	\includegraphics[width=3.2in]{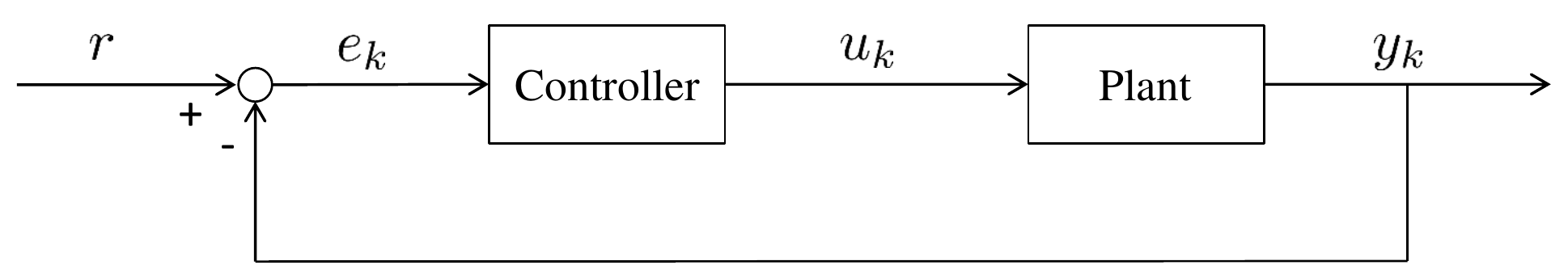}
	\caption{{\small Closed Loop Control System}}
	\label{ClosedLoopControl}
\end{figure}

The controller we choose has the form
\begin{equation}\label{eq:eq1}
u_k=u_{k-1}+A_{k}e_{k-1},\ \ \ \ \ \ \ i=1,2,\ldots,
\end{equation}
where  $A_k$  is the gain at time $k$, assumed to be positive. If $A_k=A$ where $A>0$ does not depend on time $k$ then we recognize the
controller as an adder, a discrete-time equivalent of an integrator. Since generally $A_k$ depends on $k$, we call the controller a
{\it variable-gain integrator.} The plant generally characterizes the relationship between
the control signal  $\{u_k\}$ and the output process $\{y_k\}$. Of a particular interest to us is the partial
derivative $\frac{\partial y_{k-1}}{\partial u_{k-1}}$, which we assume to be nonzero. For reasons that will become apparent
shortly,  we would like to
set the controller's gain to
$A_k=\big(\frac{\partial y_{k-1}}{\partial u_{k-1}}\big)^{-1}$. In this we assume that the partial derivative
$\frac{\partial y_{k-1}}{\partial u_{k-1}}$
is computable in real time from suitable measurements of the system, and hence can play  a  part in  the control law.
However,  such real-time computations may be subjected to delays and errors. Therefore an approximation may have to be used, resulting in the
following definition of the controller's gain,
\begin{equation}
A_k=\frac{1}{\frac{\partial y_{k-1}}{\partial u_{k-1}}+\eta_{k-1}},
\end{equation}
where $\eta_{k-1}$ denotes  an additive  error.
To complete the characterization of the loop we note that the tracking error
is
\begin{equation}
e_k=r-y_k
\end{equation}
as is evident from Figure 1. The control law consists of repeated recursive applications of Equations (2)-(1)-(3).

The rationale behind the choice of $A_{k}$ in Eq. (2) can be seen by considering  for a moment the case where the plant is a memoryless
nonlinearity, hence described by the relation
$y_k=g(u_k)$ for a differentiable function
$g:R\rightarrow R$. In this case $\frac{\partial y_{k-1}}{\partial u_{k-1}}=\frac{dg}{du}(u_{k-1})$, and Eqs. (1) - (3) result in
\begin{equation}
u_k=u_{k-1}+\frac{1}{\frac{dg}{du}(u_{k-1})+\eta_{k-1}}(r-g(u_{k-1})).
\end{equation}
We recognize this as the Newton-Raphson method for solving the equation
$r-g(u)=0$, where the derivarive term $\frac{dg}{du}(u_{k-1})$ is corrupted by  the additive error
$\eta_{k-1}$. There are well-known convergence results including a geometric convergence
rate, namely the existence of $\theta<1$ such that
\begin{equation}
|r-g(u_{k})|<\theta|r-g(u_{k-1})|,\ \ \ \ \  k=1,2,\ldots;
\end{equation}
see \cite{Lancaster66}.
In particular, Equation (5) holds under substantial upper bounds on the relative error ${\cal E}_{k-1}:=|\eta_{k-1}|/|\frac{dg}{du}(u_{k-1})|$,
hence convergence of the Newton-Raphson method is said to be robust with respect to errors in the derivative $\frac{dg}{du}(u_{k-1})$.

These results have been extended to the more-general setting where
the plant-system
is dynamic (as opposed to memoryless), stochastic and time-varying.  In such setting the term $\frac{dg}{du}(u_{k-1})$ makes no sense but the term
$\frac{\partial y_{k-1}}{\partial u_{k-1}}$ in Eq. (2) can be  well defined. One cannot expect convergence in the form of the limit
$\lim_{k\rightarrow\infty}(r-y_k)=0$ to hold true due to variations in the system's
characteristics.  However, results of the form
$limsup_{k\rightarrow\infty}|r-y_k|<\varepsilon$ were obtained in \cite{Wardi16} for a suitable $\varepsilon>0$ which depends on a measure of the system's variability.
The geometric convergence expressed in Eq. (5) is extended as long as $|r-y_k|$ is not too small, even for fairly large
errors $|\eta_{k-1}|$. These last two results imply fast approach of the tracking algorithm towards its target $r$ (though not
its convergence exactly to $r$), and its  robustness
with respect to computational errors of the derivative term $\frac{\partial y_{k-1}}{\partial u_{k-1}}$.

In the context of computer processors, this  technique was applied to regulate instructions' throughput.
The plant is modelled as a discrete event dynamic system controlled by the processor's clock rate (frequency), whose
 output is the average instruction-throughput measured over
 short time frames. The technique was verified by both simulation \cite{Almoosa12a,Chen15} and implementation on a Haswell machine \cite{Chen16}.
 In simulation  the  derivative term
 $\frac{\partial y_{k-1}}{\partial u_{k-1}}$ is estimated  by Infinitesimal  Perturbation
Analysis  \cite{Ho91,Cassandras08}, and in implementation a cruder but faster approximation is used. This paper concerns
 power regulation which poses a different set of challenges, and it uses a system characterization  as described in the following paragraphs.

The power dissipated in a core has two major components, static power and dynamic power, respectively
denoted by
$P_s$ and $P_d$; Thus
\begin{equation}
P=P_s+P_d.
\end{equation}
The dynamic power is due to the switching activities at the gates of the core. It  depends on the supply voltage $V$, clock rate
(frequency) $\phi$,
core's capacitance $C$, and the program-workload $\alpha$ representing the switching activities in the core's
transistor gates. This dependence is represented by the equation
\begin{equation}
P_d=\alpha CV^2\phi;
\end{equation}
see \cite{Hennessy12}.
Generally $C$ is a constant which can be assessed empirically, but $\alpha=\alpha(t)$ varies rapidly with the program
load and cannot be measured. The relationship between frequency and voltage often is affine, namely $V=a+m\phi$.
In this case, and in light of Eq. (7),  $P_d$ can be expressed as a third-degree
polynomial in $\phi$. However, it may be impossible to empirically determine the coefficients of this polynomial
	due to rapid variations of $\alpha(t)$. Furthermore,
	often it is impossible to measure the
	dynamic power but only the total power $P$.
	Consequently
	we are unable to compute  the coefficients of the polynomial function relating frequency to dynamic power.
	As a matter of fact, earlier attempt to apply the regulation algorithm on Haswell with various fixed polynomial coefficients
	determined off line failed to yield the desired tracking.

The static power depends on the supply voltage and temperature,
while the temperature depends on the total power  (see \cite{Hennessy12}). This circular relationship
between power and temperature precludes the existence of a simple model relating frequency to static power. Moreover,
the temperature may vary during program execution, further complicating the prospects of a frequency-to-power tractable
model that can be used in a real-time control.

As mentioned in the introduction, the  Haswell machine which serves as the implementation platform consists of four cores
processing eight concurrent threads. All of the cores reside in the same voltage island,
hence we cannot  control each one of them  separately  but rather control them jointly by a common frequency,
the {\it processor frequency}. The controlled quantity is the average power among the four cores, called
the {\it processor power}. In the setting of Figure 1, we partition the time horizon into equally-spaced contiguous intervals
called {\it control cycles} and denoted by $C_{k}$, $k=1,2,\ldots$; $u_{k}$ is the processor frequency applied during
$C_{k}$, and $y_{k}$ is the average among the cores  of the mean  spatial and temporal  power measured during $C_k$ at each core. A key
question is how to obtain an estimate of the derivative term
$\frac{\partial y_{k-1}}{\partial u_{k-1}}$, in Equation (2). This requires knowledge of some parameters of the
plant model relating $u_k$ to $y_k$, but  such a model is only partly available. In fact, we mentioned that
there in no analytic model for the static power, and while there is an adequate  third-degree polynomial for the dynamic power, its coefficients change with time at a high rate.

We overcome these problems by the following approach. First, we search for a third-order polynomial for estimating the relation between the applied
frequency and the total processor power. This of course can fit the dynamic power but not the static power.
However, in computing applications
at the frequency range considered in this paper, the static power comprises 20\% - 30\% of the total power, and therefore we feel
confident leveraging the aforementioned robustness of the performance of the tracking controller with respect to errors in computing
$\frac{\partial y_{k-1}}{\partial u_{k-1}}$.
Second, to cope with the rapid variations in the coefficients of this polynomial due to changes in $\alpha(t)$, we use
a system identification module run concurrently, in real time, along the program-executions by the processor.
	
Let us denote by $p_k(\phi):=a_k\phi^3+b_k\phi^2+c_k\phi+d_k$ the estimator polynomial during $C_{k}$,
then its derivative $\frac{dp_{k-1}}{d\phi}(\phi_{k-1})=3a_{k-1}\phi_{k-1}^2+2b_{k-1}\phi_{k-1}+c_{k-1}$ is the term
$\frac{\partial y_{k-1}}{\partial u_{k-1}}+\eta_{k-1}$ in Equation (2).
A system identification module, comprised of a standard recursive least-square estimator (e.g., \cite{Keesman11}),  is used to
compute  the coefficient-vector $x_k:=(a_k,b_k,c_k,d_k)$ during $C_{k-1}$.

It must be pointed out that  performance of the power regulator is affected by several practical considerations. First, the rate at which energy and power can be measured is determined by the processor vendor, which in this case is Intel. The model specific registers are updated at approximately 1 ms intervals but no timestamp is provided so it is not possible to know when the measurement interval began. Consequently, mapping measurements to application code is difficult and can cause larger deviations in regulated power than the model would otherwise achieve. Second, the current manner in which the frequency is changed is via file I/O incurring substantial latency relative to the execution time of instructions. If program behavior changes significantly during this interval, tracking becomes more challenging. For example, data center programs that possess poor spatial and temporal reference locality and are memory intensive will exhibit wide variations in average instruction execution time due to memory accesses. High latency in setting the processor frequency will make it difficult to rapidly adapt to changes in power consumption and consequently will affect the rate of convergence of the power regulator and the choice of the duration of the control cycle. Such practical considerations must be overcome by robustness in the design of the regulator.

\section{Results}

The proposed power regulator was tested on various programs from two suites of industry benchmarks, Splash 2 and GraphBig. Splash 2 is a set of standard benchmark programs for shared memory cache coherent multiprocessors \cite{Woo95}. It includes a collection of multithreaded workloads representing traditional engineering and science applications. The majority of the benchmarks are from the traditional high performance computing domain while several are drawn from signal processing and general engineering computations such as computer graphics. GraphBIG is a set of benchmark programs that perform computations over graphs \cite{Nai15},
and was inspired by the IBM System G project,
surveyed in the sequel.

We are witnessing  an explosive growth in modern data science applications executing in data centers that deal with data that is of the relational form and can be represented by graph data structures with large numbers of node and edge properties. These applications have irregular memory access patterns, exhibit low spatial and temporal locality, and are characterized by low operation density, i.e., number of operations per byte of data accessed. Consequently, they stress the memory system and challenge optimizations for achieving high processor utilization. To cover major graph computation types and data sources encountered in such data center applications, GraphBIG incorporates representative data structures, workloads and data sets from 21 real-world use cases from multiple application domains. As a complement to traditional science and engineering applications represented by the Splash-2 benchmarks, GraphBIG represents the relational computation driving commercial sectors such as retail forecasting, data analytics, finance, and banking.

We implemented the controller by loading a C++ program
to the Haswell processor via the PAPI interface \cite{Browne00}.\footnote{Modern
microprocessors include  hardware counters that record the occurrences of various events during program executions,
	like completion of instructions' executions, cache misses, etc.    The Performance Application Programming Interface (PAPI) is a publicly available software infrastructure for accessing these performance counters during program execution.}
The Haswell processor  has a finite set of
16 frequencies, namely $\Omega := \{0.8,1.0,1.1,1.3,1.5,1.7,1.8,2.0,2.2,2.4,2.5,2.7,2.9,3.1,\\
3.2,3.4\}$ in GHz.
Therefore, we augmented Eq. (1) by projecting its Reft-Hand Side (RHS) onto $\Omega$. That is,
with $P_{\Omega}(u):={\rm argmin}\{|v-u|:v\in\Omega\}$ for $u\in R$ (with $v<u$  if the argmin is not unique),
we replace (1) by the following equation,
\begin{equation}
u_{k}=P_{\Omega}(u_{k-1}+A_ke_{k-1}).
\end{equation}
The control algorithm consists of a recursive application of Eqs. (2)-(8)-(3).

The control cycles of the algorithm can be chosen according to  performance considerations such as settling times, subject to
hardware constraints.   At the end of each control cycle, first the model parameters are recomputed  by the system identification module and then
the operating frequency is assigned to the processor for the next period.
In the Haswell
processor that we use, energy consumption values are provided at a sampling interval of 1 ms. Hence we pick control cycles that are multiples of this interval.\footnote{The power measurements are indirect. Haswell
	provides  an energy counter, so we can measure the energy in 1 ms  and divide the result by the measurement time  to obtain the average
	power during $1~ms$ periods.}
 We test the control algorithm at two different rates associated with
the control cycles of  $10$ ms and $30$ ms, and  we depict
graphs of power  as function of time during the first $4,000~ms$ of program executions.

\subsection{Splash-2 benchmark programs}
We tested the controller on two programs: {\it Barnes}, and {\it Ocean-nc}.
Barnes is a compute intensive application with under  $10\%$  of memory-access instructions. In contrast,
Ocean-nc is a memory-intensive program with memory-bound instructions
in the $30\%-50\%$ range for typical applications.
In both cases we set the power-target value to $10~W$.

Consider first Barnes. For control cycles of $10~ms$,
results of an application of the control
algorithm are shown in Figure 2 and Figure 3.  Figure 2 depicts the graph of power vs. time for the fist $4,000~ms$, corresponding to
$400$ iterations.  The power rises from an initial value of $6.34$ W, and following a period of transient behavior lasting $720~ms$, taking 72 control cycles,
it settles into an oscillatory behavior about the target value of 10 W.
The average power computed over the interval
$[720,4000]~ms$ (after the power has settled for the first time into an oscillation about the target level)
 is $10.2604$ W, which is $0.2604$ W above the target level of 10 W.

 The graph of the frequency (clock rate) vs. time is  shown in Figure~\ref{fig:FreqBarnesTarget10Cycle10}.
 The larger oscillations in the first $1,400~ms$ likely are due to the fact that
early in the program there are more memory instructions than in the last $2,600~ms$.
Memory instructions can take one-to-two orders more time than computational instructions. Therefore memory-intensive periods tend to have  greater
variability in  the program workload and hence  larger changes in frequency and power.
The persistence of the smaller oscillations throughout the interval $[1400,4000]~ms$ likely is due to
quantization effects associated with the fact that the frequency-set $\Omega$ is finite.
The average frequency in the interval
$[720-4000]~ms$ is $1.93$ GHz.

\begin{figure}	[!t]
	\centering
	\includegraphics[width=3.2in]{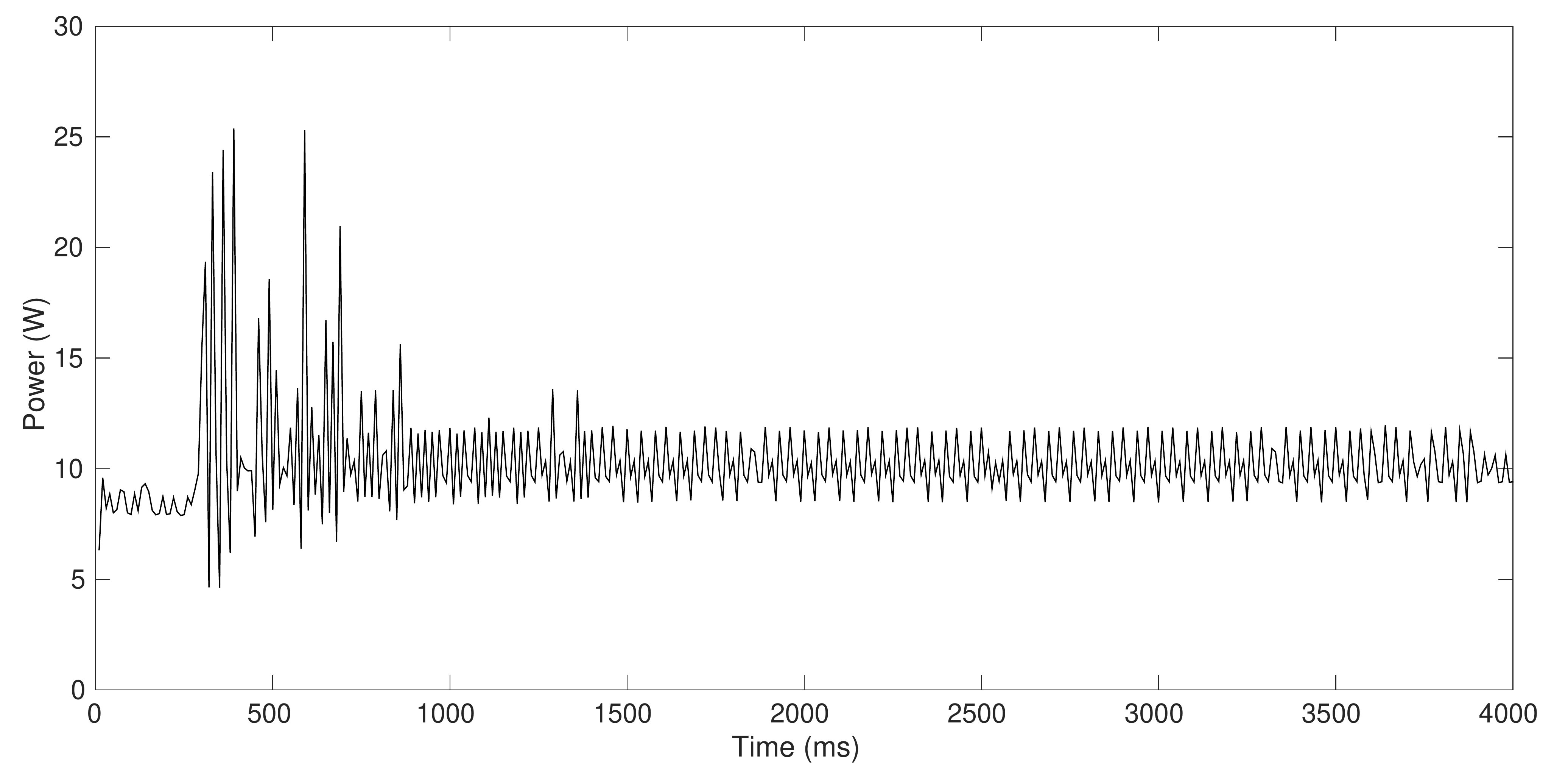}
	\caption{{\small Barnes: power vs. time,  control cycle = 10 ms}}
	\label{fig:PowerBarnesTarget10Cycle10}
\end{figure}

\begin{figure}	[!t]
	\centering
\includegraphics[width=3.5in]{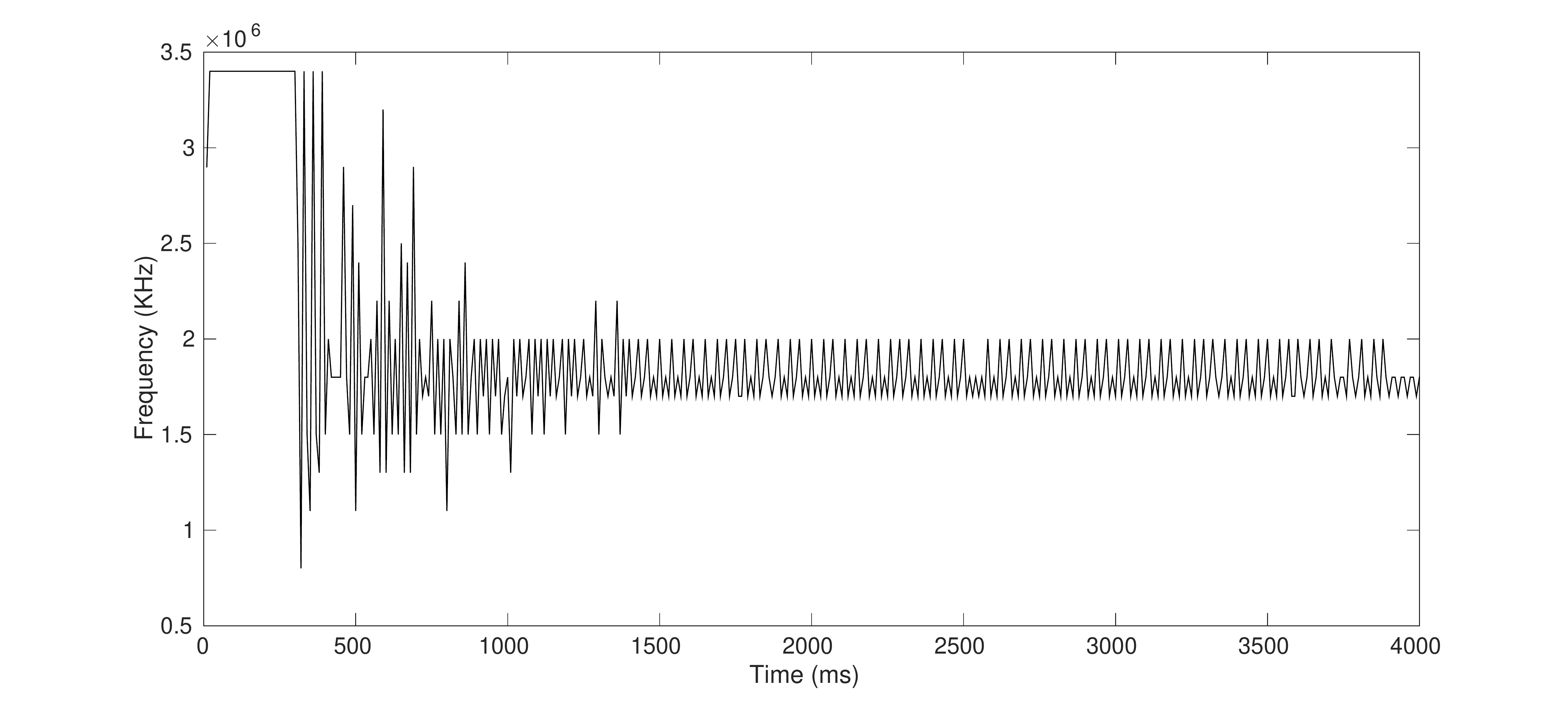}
	\caption{{\small Barnes: clock frequency vs. time,  control cycle = 10 ms}}
	\label{fig:FreqBarnesTarget10Cycle10}
\end{figure}

\begin{figure}	[!t]
	\centering
	\includegraphics[width=3.2in]{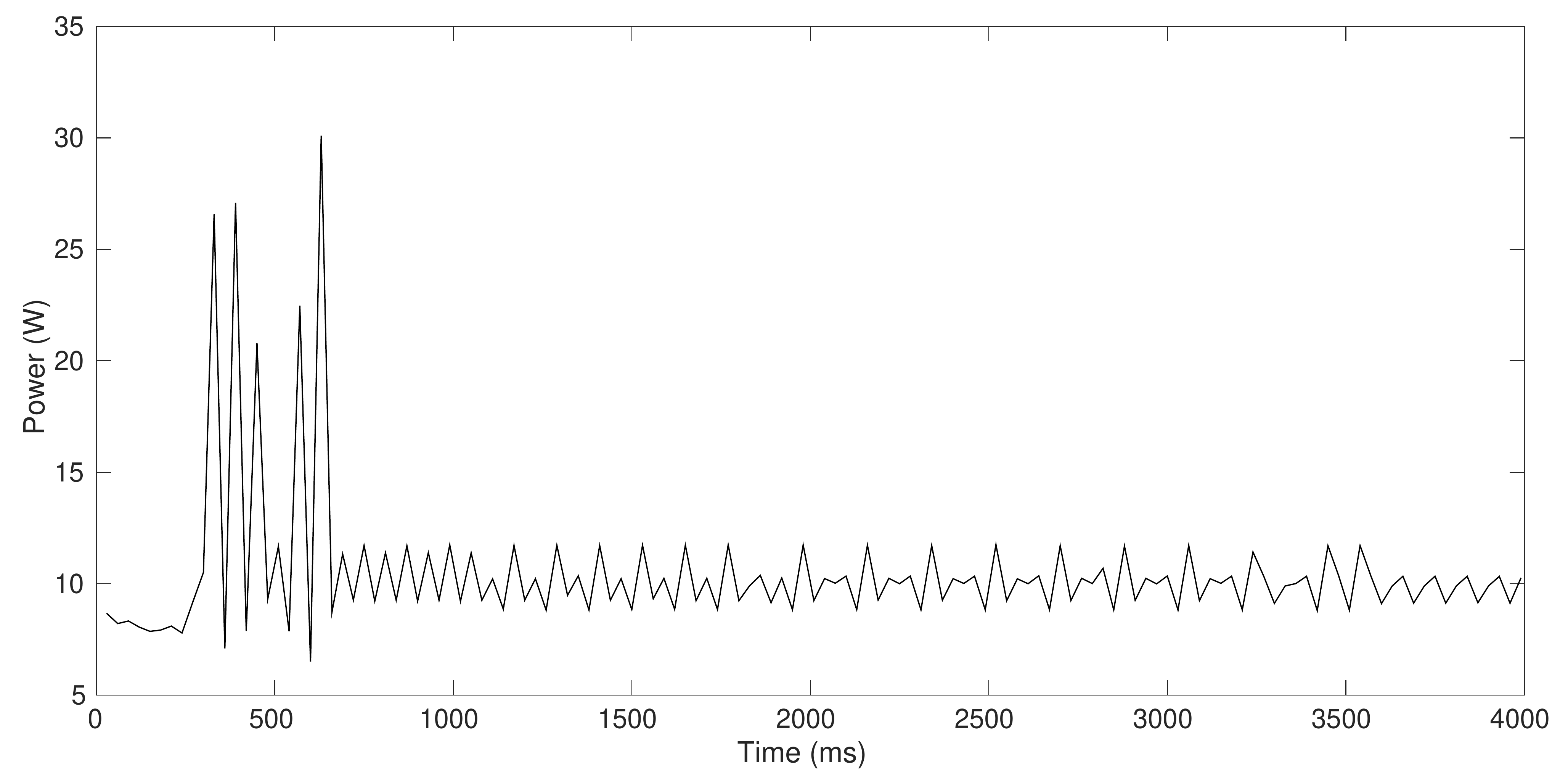}
	\caption{{\small Barnes: power vs. time,  control cycle = 30 ms}}
	\label{fig:PowerBarnesTarget10Cycle30}
\end{figure}

For the control cycle of $30~ms$,   the graph of power vs. time is depicted in  in Figure~\ref{fig:PowerBarnesTarget10Cycle30}.
The power rises from an initial value of $8.6543$ W, and after 720 ms (or 24 control cycles) it settles around the target value of 10 W. Its average
in the interval $[720,4000]~ms$  is $10.4344$ W, which is $0.4344$ W over the target level of 10 W. The frequency profile is similar to that of Figure 3 and hence not shown, and its average in the interval
$[720,4000]~ms$ is  $1.89$ GHz.

Table I  summarizes the time it takes the power to settle about its target value for the first time, as well as the absolute value of
the error (difference)
between the average power and the target value of $10~W$.  The performance of the controller is similar for the two
control cycles, and the indicated minor differences likely are due to the frequency quantization.

\begin{table}[]
	\centering
	\caption{Barnes: power error and settling time}
	\label{tableBarnes}
	\begin{tabular}{|l|l|l|l|}
		\hline
		Control Cycle (ms) & 10  & 30   \\ \hline \hline
		Error (W)           & 0.2604  & 0.4344  \\ \hline
		Settling Time (ms)      & 720   & 720 \\ \hline
	\end{tabular}
\end{table}

\begin{figure}	[!t]
	\centering
	\includegraphics[width=3.2in]{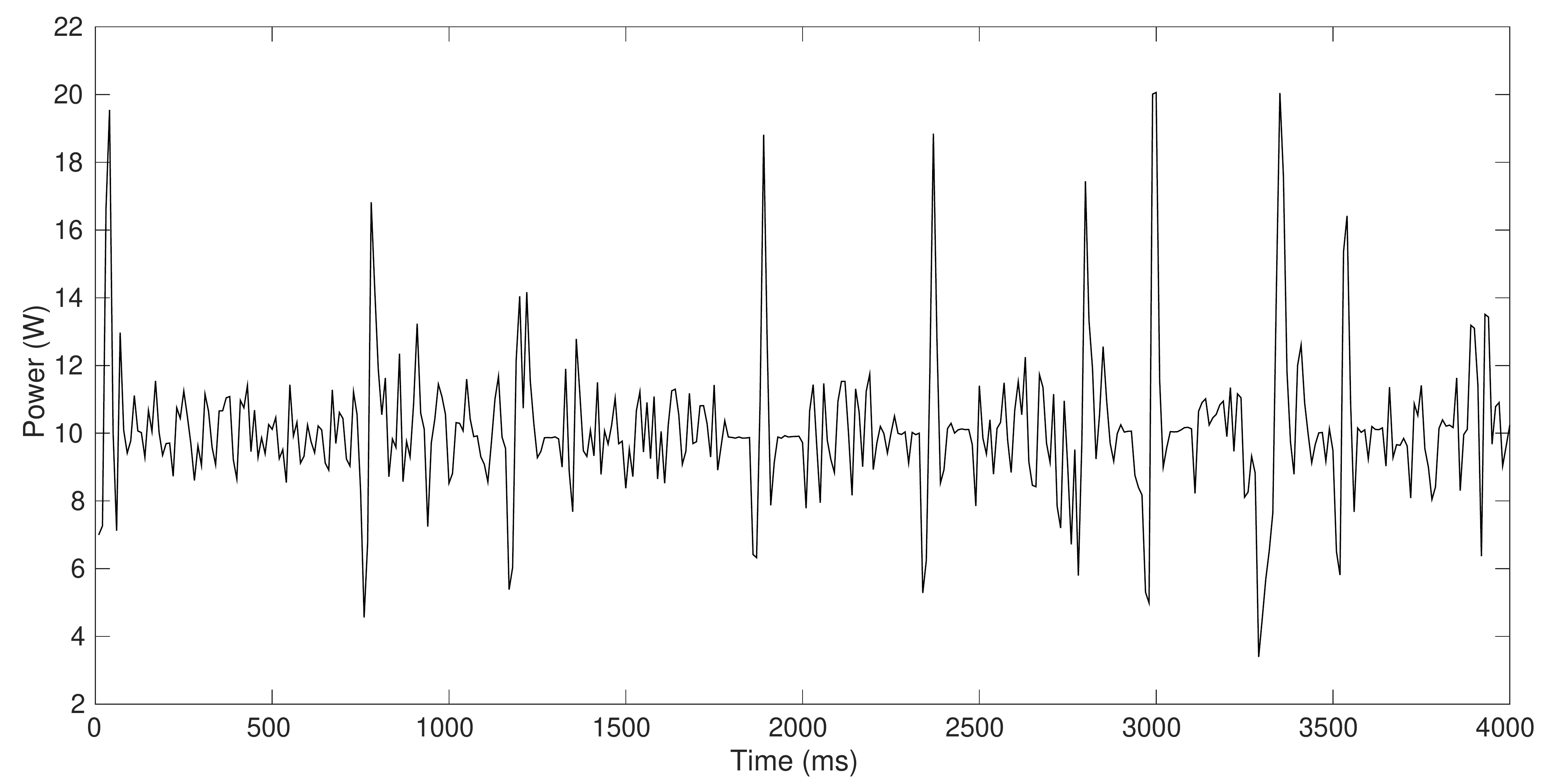}
	\caption{{\small Ocean-nc: power vs. time,  control cycle = 10 ms}}
	\label{fig:PowerOceanTarget10Cycle10}
\end{figure}

\begin{figure}	[!t]
	\centering
	\includegraphics[width=3.2in]{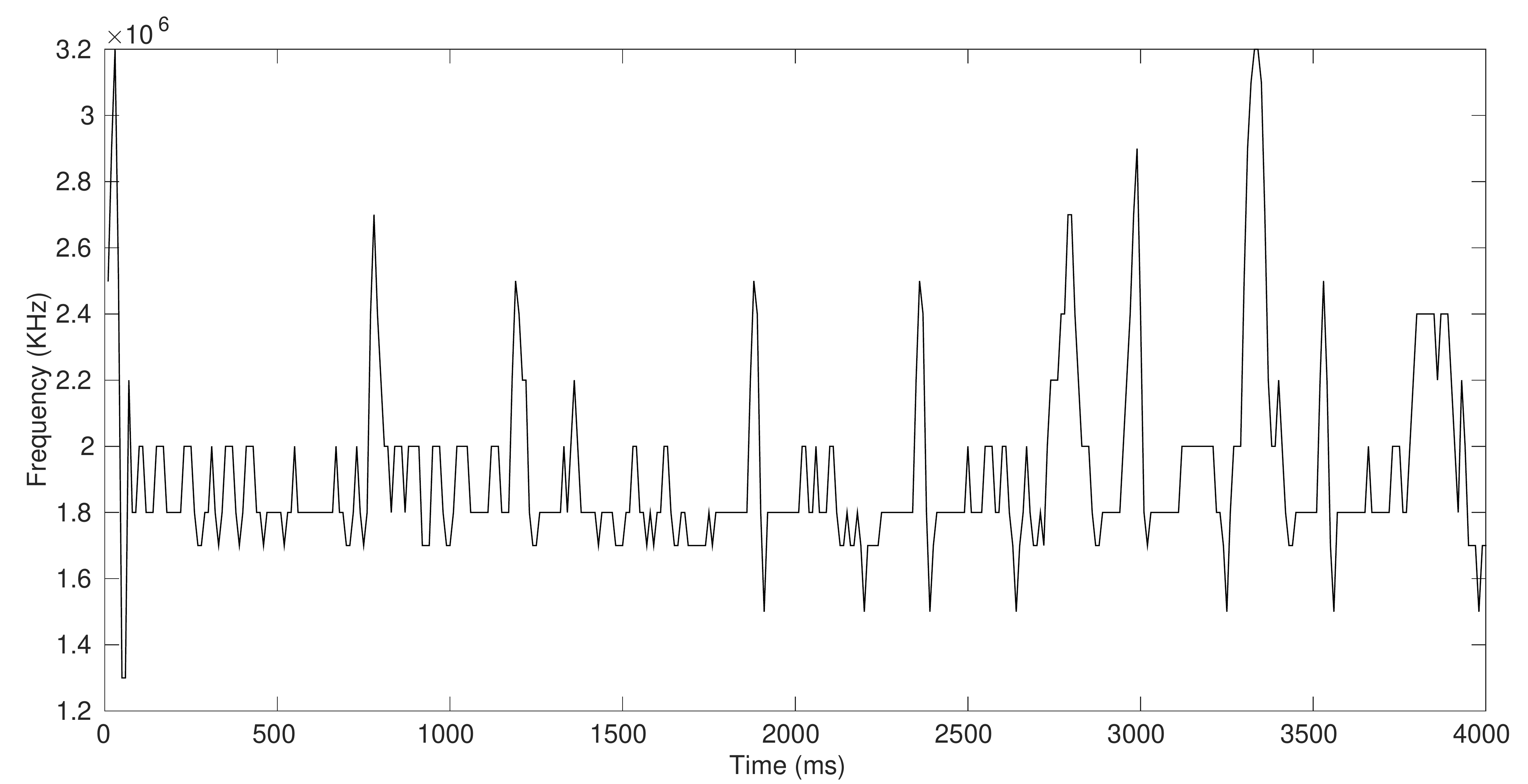}
	\caption{{\small Ocean-nc: clock frequency vs. time,  control cycle = 10 ms}}
	\label{fig:FreqOceanTarget10Cycle10}
\end{figure}

Consider next Ocean-nc.   Graphs of power and frequency vs. time are depicted in  Figure~\ref{fig:PowerOceanTarget10Cycle10}
and Figure~\ref{fig:FreqOceanTarget10Cycle10}, respectively.  The power rises from an initial value of $7.015$ W, and following a period of transient behavior lasting $1,240~ms$, taking 124 control cycles, it settles into an oscillatory behavior about the target value of 10 W.
The average power computed over the interval $[1240,4000]~ms$ is $10.1269$ W, which is $0.1269$ W above the target level of 10 W. The average frequency is $1.918$ GHz.

\begin{figure}	[!t]
	\centering
	\includegraphics[width=3.2in]{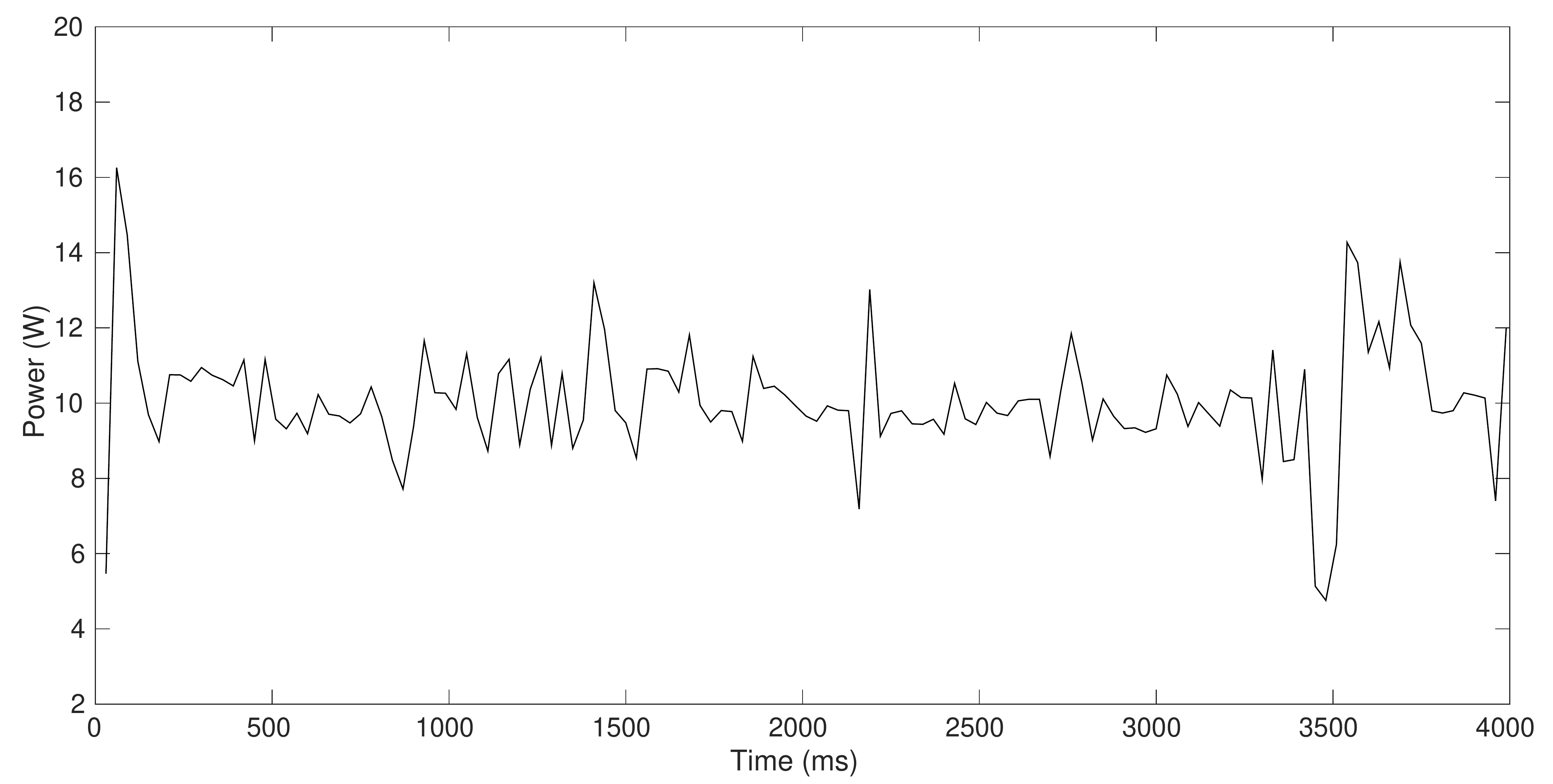}
	\caption{{\small Ocean-nc: power vs. time,  control cycle = 30 ms}}
	\label{fig:PowerOceanTarget10Cycle30}
\end{figure}

For $30~ms$-control cycles,   the graph of power vs. time is depicted in  in Figure~\ref{fig:PowerOceanTarget10Cycle30}
while the frequency graph displays similar characteristics to that in Figure \ref{fig:FreqOceanTarget10Cycle10} for
$10~ms$-control cycles, hence not shown.
The power rises from an initial value of $5.483$ W, and after $1,710~ms$ (or 57 control cycles) it settles around the target value of 10 W. Its average
in the interval [1710,4000] ms  is $9.9298$ W, which is $0.0702$ W below the target level of 10 W.  The average frequency is $1.86$ GHz. The results are summarized in Table II. A comparison between the results for Barnes and Ocean-nc will be made in
Subsection III.C, below.

\begin{table}[]
	\centering
	\caption{Ocean-nc: power error and settling time}
	\label{tableOcean}
	\begin{tabular}{|l|l|l|l|}
		\hline
		Control Cycle (ms) & 10  & 30   \\ \hline \hline
		Error (W)           & 0.1269 &  0.0702  \\ \hline
		Settling Time (ms)      & 1240    & 1710 \\ \hline
	\end{tabular}
\end{table}

\subsection{GraphBig benchmark experiments}
We tested the controller on two GraphBig programs: {\it Breadth-first Search (BFS)}, and {\it Kcore}.
BFS
 is one of the most fundamental
operations of graph computing, while kCore  encompasses topological analysis of graphs and is representative of approaches to the structural analysis of graphs. Both programs represent large-scale computations executed over
clusters of servers in large data centers.

For both BFS and Kcore the target power is $5~W$. The reason it is less that the target for the Splash-2 programs ($10~W$)
is that a
GraphBig program typically has a higher fraction
 of memory-access instructions than Splash-2 programs, which tend to be low-frequency, low-power
operations.

\begin{figure}	[!t]
	\centering
	\includegraphics[width=3.2in]{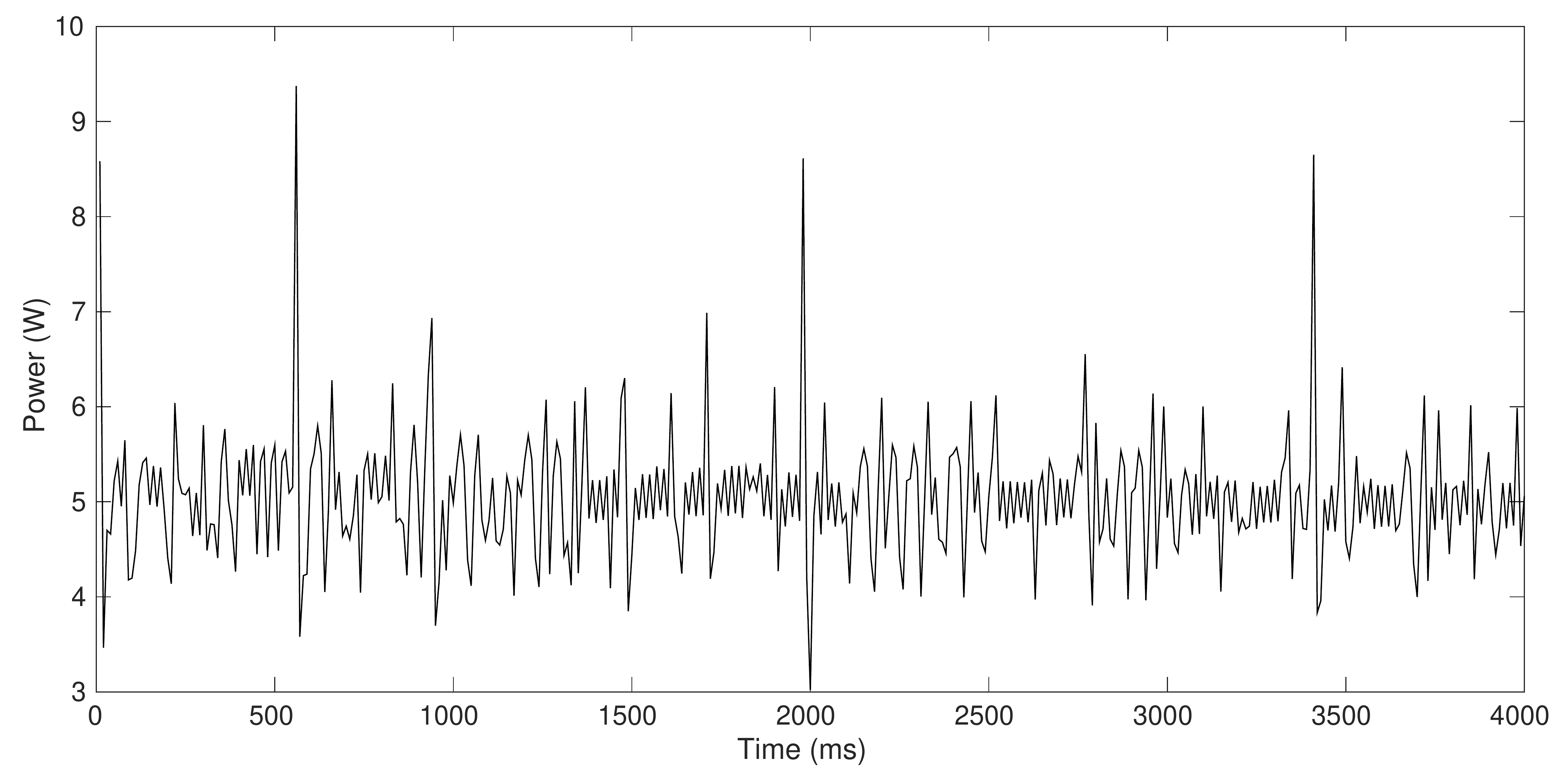}
	\caption{{\small BFS: power vs. time,  control cycle = 10 ms}}
	\label{fig:PowerTCTarget5Cycle10}
\end{figure}

\begin{figure}	[!t]
	\centering
	\includegraphics[width=3.2in]{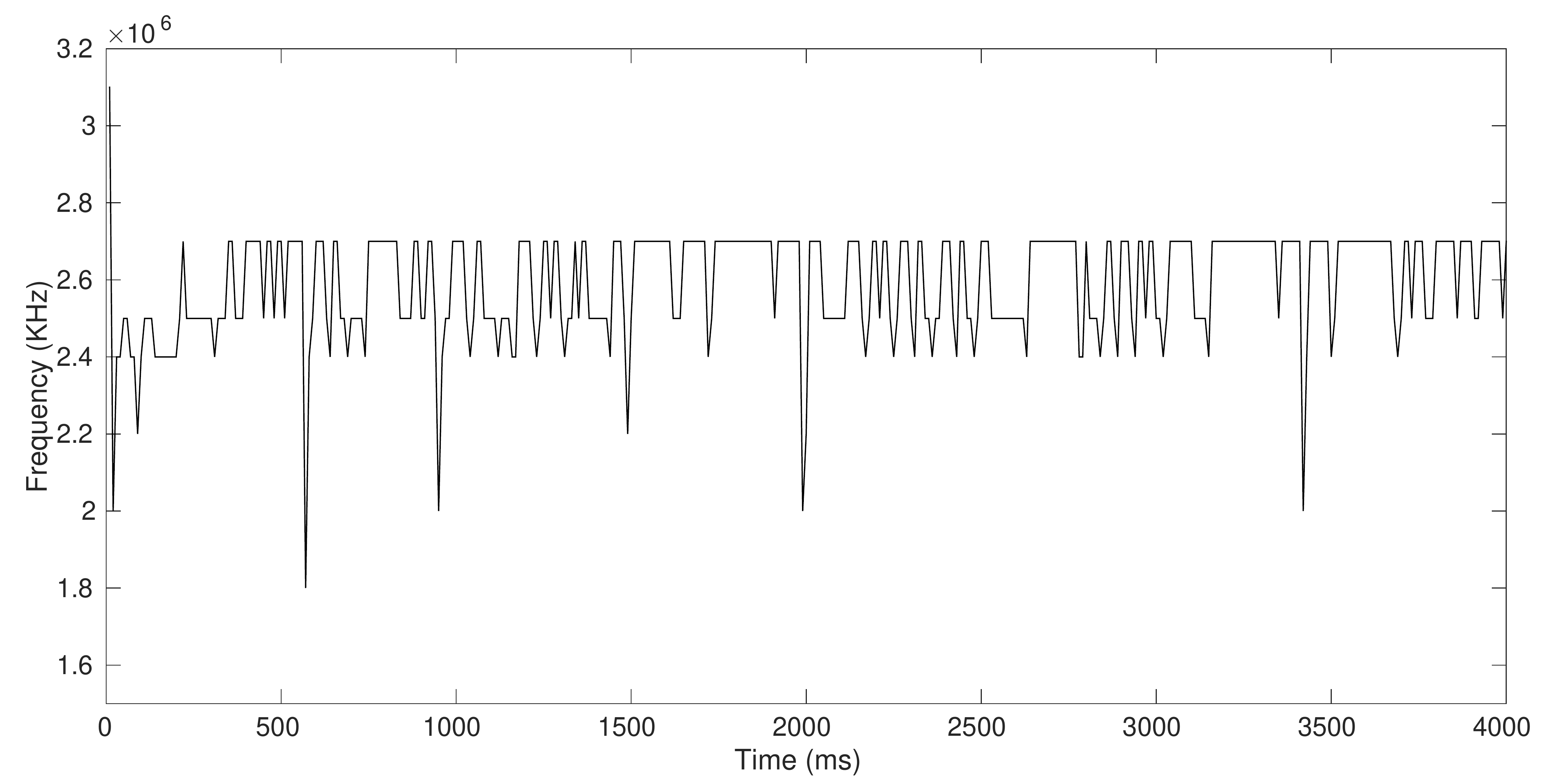}
	\caption{{\small BFS: clock frequency vs. time,  control cycle = 10 ms}}
	\label{fig:FreqTCTarget5Cycle10}
\end{figure}

Consider first the BFS program.
For  $10~ms$-control cycles, the results are shown in Figure~\ref{fig:PowerTCTarget5Cycle10} and Figure~\ref{fig:FreqTCTarget5Cycle10}. The power vs. time graph is depicted in Figure~\ref{fig:PowerTCTarget5Cycle10}.
The power starts at the  initial value of
$8.57$ W, and following an initial transient lasting 380 ms (or 38 control cycles) it settles about the target value of 5 W.  The average power
in the interval $[380,4000]~ms$  is $5.0542$ W, which is $0.0542$ W more than the target level of $5$ W.
The frequency graph is depicted  in Figure~\ref{fig:FreqTCTarget5Cycle10}, and the average frequency
in the interval $[380,4000]~ms$ is $2.59$ GHz.

\begin{figure}	[!t]
	\centering
	\includegraphics[width=3.2in]{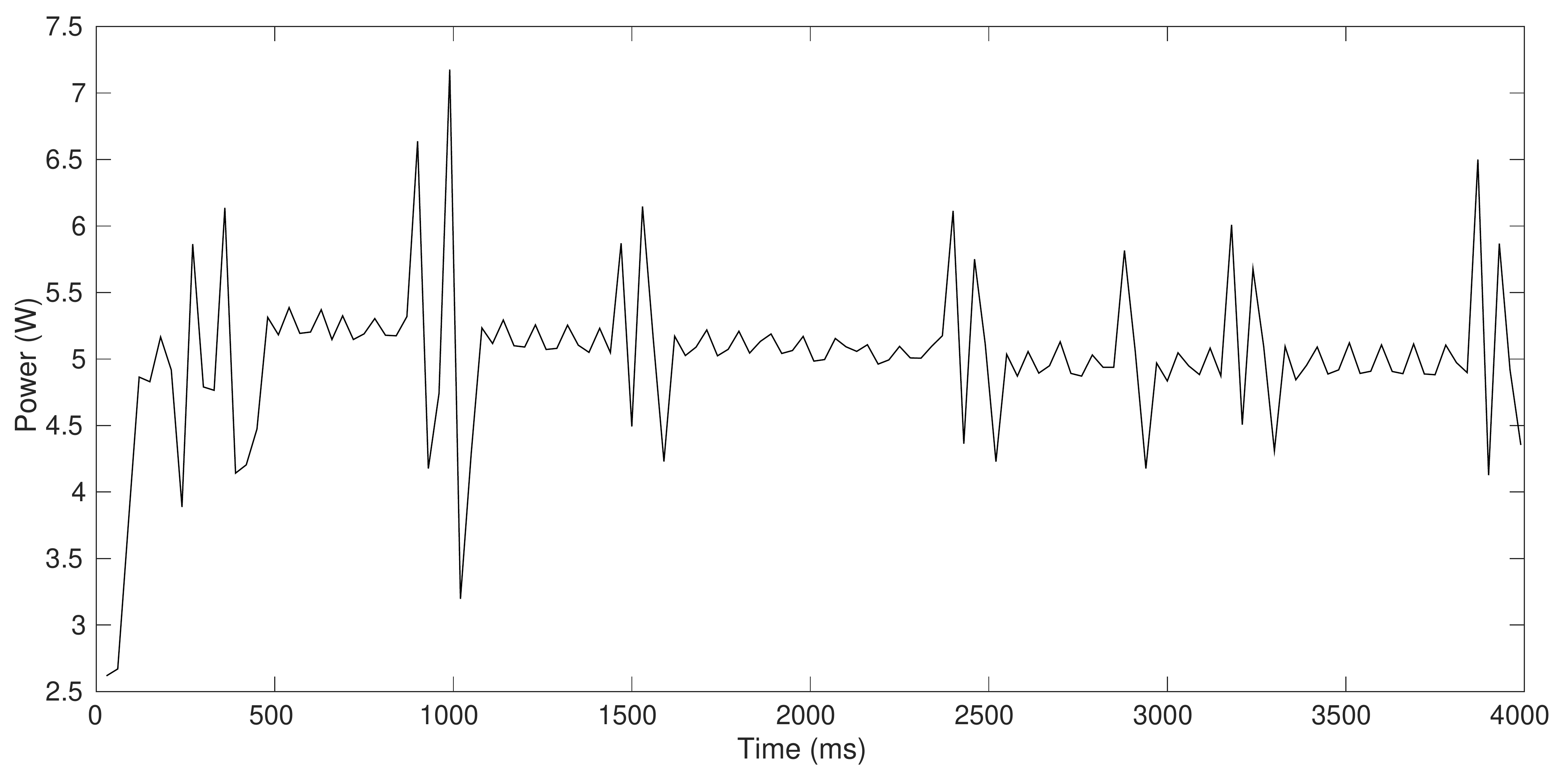}
	\caption{{\small BFS: power vs. time,  control cycle = 30 ms}}
	\label{fig:PowerTCTarget5Cycle30}
\end{figure}

For  $30~ms$-control cycles, the power graph is  shown in Figure~\ref{fig:PowerTCTarget5Cycle30}; the frequency-graph displays
similar characteristics to that for $10~ms$-control cycles depicted in Figure 9,   hence not shown. The power starts at the value of $2.62$ W, and after a transient period
of  $510$ ms (or $17$ control cycles), it settles in a band around 5 W. Its   average in the interval $[510,4000]~ms$  is
$5.0108$ W, which is $0.0108$ W more than the target level of $5$ W.
The settling time and error for BFS  for the two control cycles  are  shown in Table III.

 \begin{table}[]
	\centering
	\caption{BFS: power error and settling time}
	\label{tableTC}
	\begin{tabular}{|l|l|l|}
		\hline
		Control Cycle (ms) & 10  & 30   \\ \hline \hline
		Error (W)           & 0.0542  & 0.0108  \\ \hline
		Settling Time (ms)  & 380 & 510 \\ \hline
	\end{tabular}
\end{table}

\begin{figure}	[!t]
	\centering
	\includegraphics[width=3.2in]{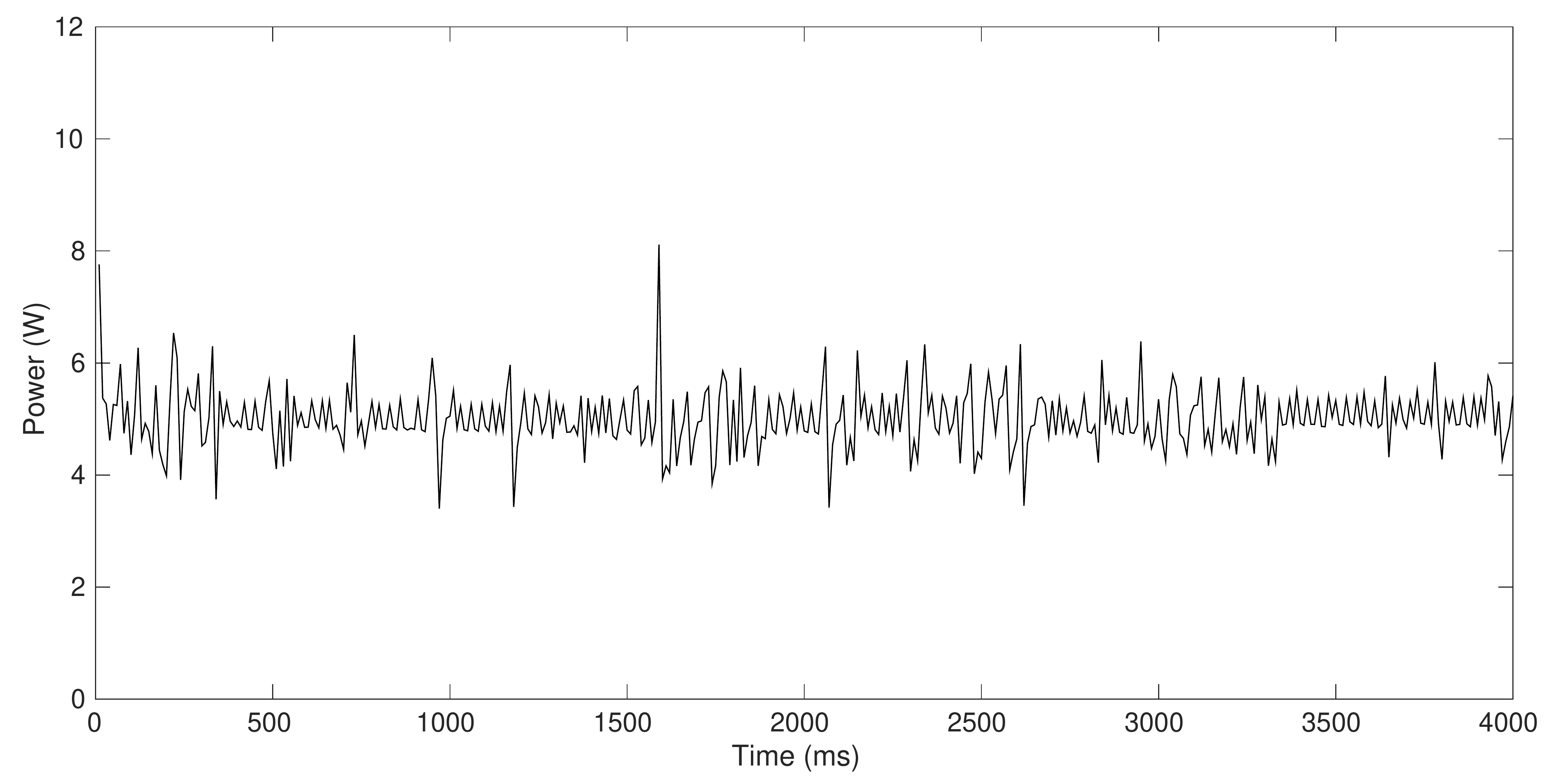}
	\caption{{\small KCore: power vs. time,  control cycle = 10 ms}}
	\label{fig:PowerKCoreTarget5Cycle10}
\end{figure}

\begin{figure}	[!t]
	\centering
	\includegraphics[width=3.2in]{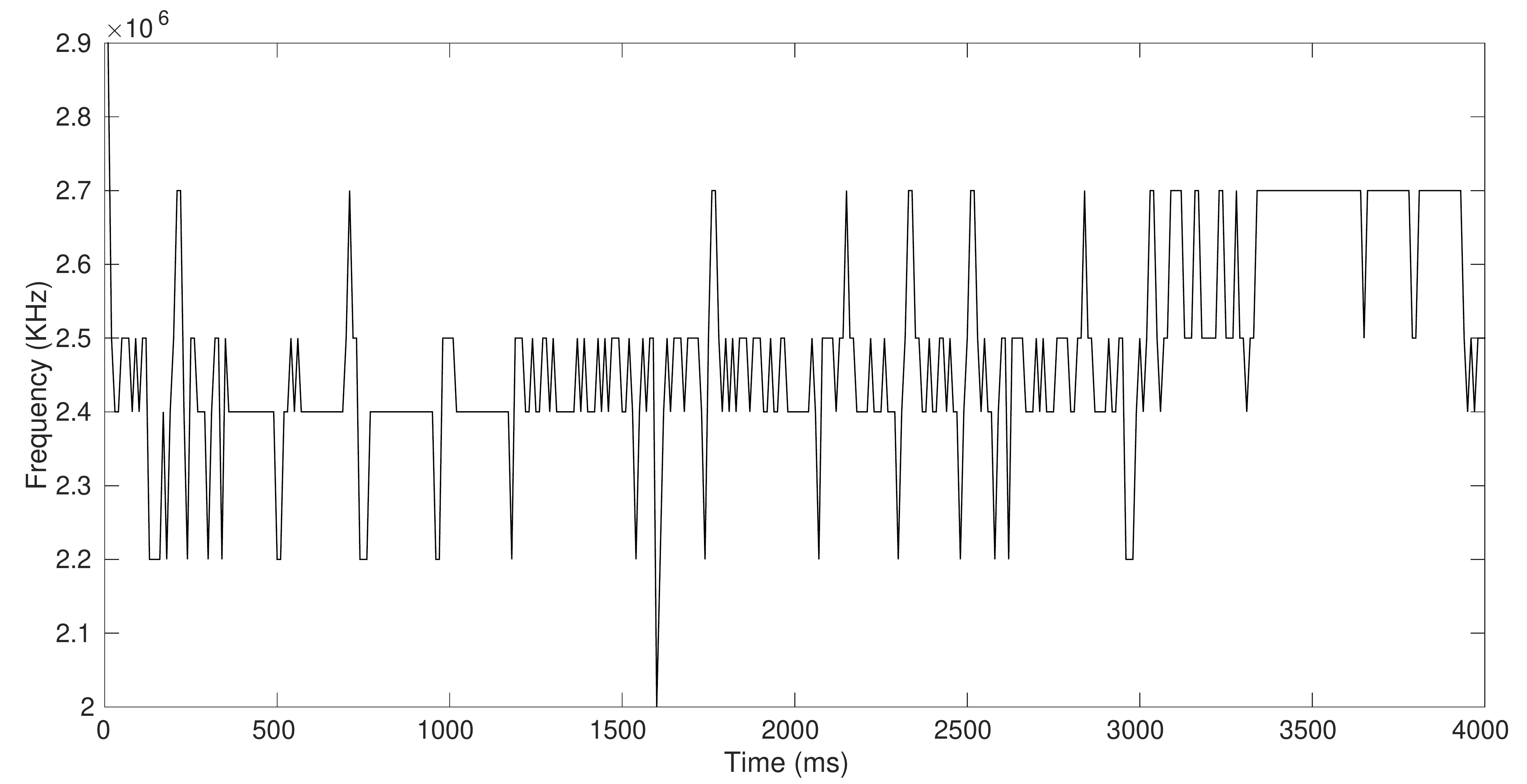}
	\caption{{\small KCore: clock frequency vs. time,  control cycle = 10 ms}}
	\label{fig:FreqKCoreTarget5Cycle10}
\end{figure}

Consider next the results for Kcore.
For a $10~ms$-control cycle, the graphs of power and frequency vs. time  are shown in Figure~\ref{fig:PowerKCoreTarget5Cycle10} and Figure~
\ref{fig:FreqKCoreTarget5Cycle10}, respectively.
The power starts at the initial value of
$7.749$ W, and following an initial transient lasting 400 ms (or 40 control cycles) it settles about the target value of 5 W.  The average power
in the interval [400,4000] ms is $5.0124$ W, which is $0.0124$ W more than the target level of $5$ W.
The frequency graph is depicted in Figure~\ref{fig:FreqKCoreTarget5Cycle10}, and its mean  is $2.478$ GHz.

\begin{figure}	[!t]
	\centering
	\includegraphics[width=3.2in]{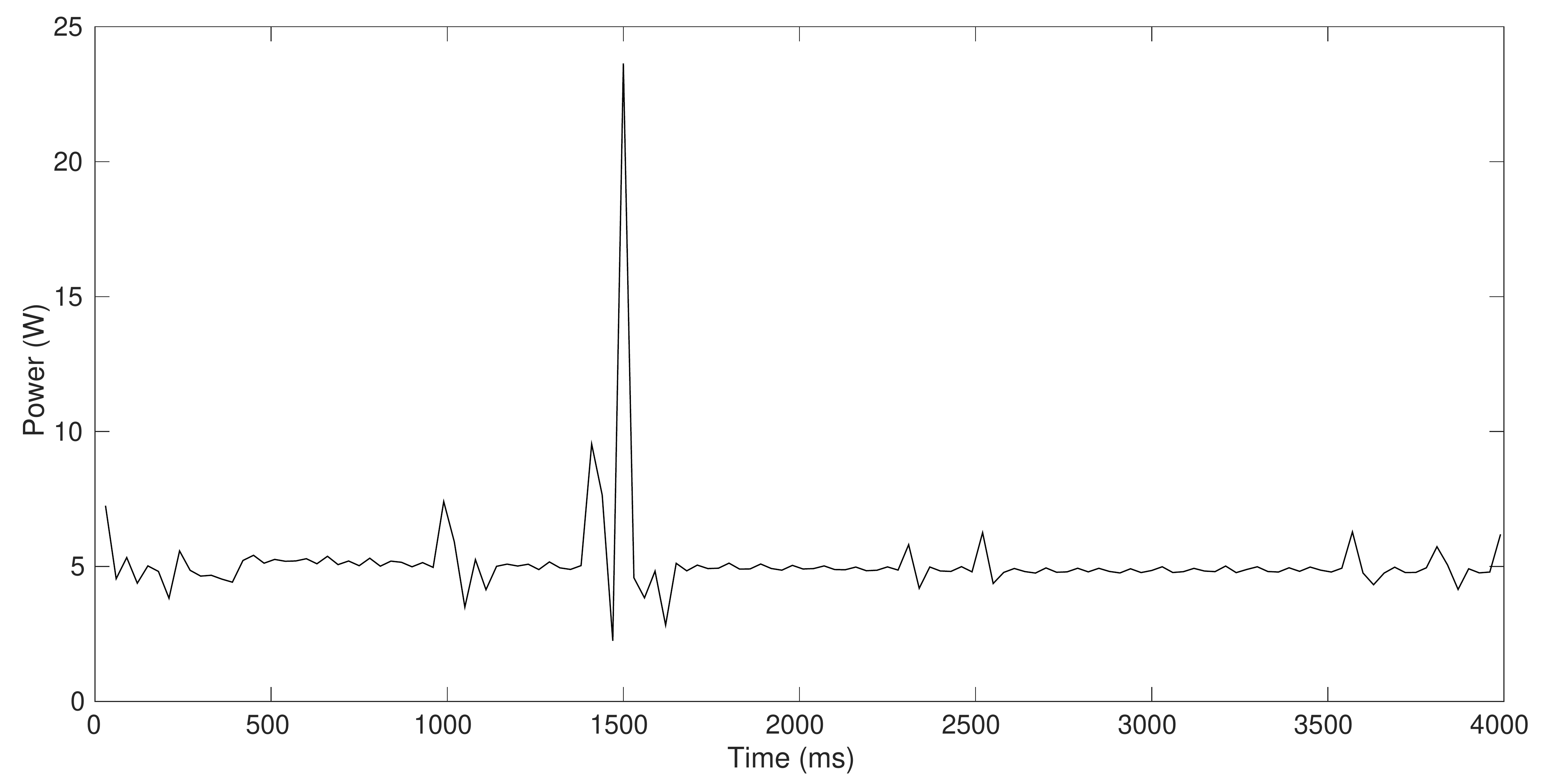}
	\caption{{\small KCore: power vs. time,  control cycle = 30 ms}}
	\label{fig:PowerKCoreTarget5Cycle30}
\end{figure}

For  $30~ms$ control cycles, the power graphs are shown in Figure~\ref{fig:PowerKCoreTarget5Cycle30}. The power starts at the value of $7.23$ W, and after a transient period
of  $480$ ms (or $16$ control cycles), it settles in a band around 5 W. Its   average in the interval [480,4000] ms is
$5.1291$ W, which is $0.1291$ MIPS more than the target level of $5$ W. The frequency graph is not shown since it is similar to the
graph depicted in Figure 12 for the case of $10~ms$.  The average frequency is $2.608$ GHz. These results are shown in Table  \ref{tableTC}.

\begin{table}[]
	\centering
	\caption{KCore: power error and settling time}
	\label{tableTC}
	\begin{tabular}{|l|l|l|}
		\hline
		Control Cycle (ms) & 10  & 30   \\ \hline \hline
		Error (W)           & 0.0124  & 0.1291  \\ \hline
		Settling Time (ms)  & 400  & 480 \\ \hline
	\end{tabular}
\end{table}

\subsection{Discussion and comparison of results}

The entries in the four tables, namely average error and settling
times, capture the performance of the regulation algorithm as applied
to the four respective programs. The differences in behavior arise
mainly from differences in the compute vs. memory behavior of the
programs. Recall that the regulation algorithm is being applied to the
operation of the cores on the processor chip - the major source of
power dissipation today. The memory system is off-chip and operates in
a different voltage island and off a different clock. Specifically,
the processor power is directly controlled by the processor
clock-rate, but only indirectly by the impact of memory
instructions. Memory instructions can take two orders-of-magnitude
more time to complete execution than compute instructions. Therefore
the cores can stall for periods of time while waiting for memory
access operations to complete. During stalls the processor consumes
less power.

Further, consider the execution of a program that is memory-bound,
i.e., its execution time is determined by how fast memory references
can be satisfied. In this case, running the cores in the processor at
a higher speed consumes more power but produces no appreciable
improvement in execution time. Conversely, a compute bound application
consumes more power and takes less time when core frequency is
increased. Furthermore, execution time of memory instructions can be
highly variable due to congestion on the memory bus and queuing delays
in the memory system. These  observations help explain some of
the results of the various experiments described in the previous
subsections.  We must point out, however, that we have a limited
number of experiments for any sweeping conclusions, and those may have
to wait for larger volume of data to be collected in the future.

Consider first a comparison between the two Splash-2 programs, Barnes
and Ocean-nc. Barnes is compute intensive and Ocean-nc is memory
intensive.  Therefore we expect the power graph of Ocean-nc to display
larger variability than the power graph of Barnes. That indeed can be
seen by comparing the respective graphs in Figure 2 and Figure 5 for
$10~ms$ control cycles, and in Figure 4 and Figure 7 for $30~ms$
control cycles.  This also explains the larger settling time of
Ocean-nc vs. Barnes, as can be seen in Table I and Table II. On the
other hand, the tables show that the error measure for Barnes is
larger than for Ocean-nc. We explain this by noting that the frequency
fluctuations of Barnes are smaller than those of Ocean-nc, as expected
and also shown in Figure 3 and Figure 6. If the frequency set $\Omega$
were continuous then we would expect Barnes to have the smaller
error. However, the fact that $\Omega$ is a finite set suggests, by
Eq. (8), that  the frequency   can get trapped at a particular
value which leads to quantization. It is possible that the wider
frequency variations of Ocean-nc makes it easier for the frequency to
escape from a given value thereby reducing the quantization and
resulting in smaller error.

Comparing results of the Splash-2 programs to those obtained from the
GraphBig programs, the main difference is in the error measures.
According to Tables I-IV, the errors for BFS and Kcore are smaller
than for Barnes and Ocean-nc.  Like Ocean-nc, both BFS and Kcore are
also sensitive to the performance of the memory system, although for
different reasons than Ocean-nc. These applications process large
graphs. The ratio of compute instructions to memory instructions is
smaller and the patterns of memory  references are quite irregular -
this makes the memory behaviors more sensitive to memory system
latency. This can cause more frequent stalls by the processor
cores. The regulation algorithm takes one step per control cycle and
control cycles are independent of the number and duration of stalls.
Therefore, on average there are less computing activities per control
cycle in BFS and Kcore than in Ocean-nc; also less than Barnes which
is compute intensive. This results in longer averages per frequency-variable which in turn
translates into more precise computations.

Finally, we mention that for each given program there is no noticeable
difference in performance between the experiments with the respective
control cycles of $10~ms$ and $30~ms$. We also note that performance
of the regulation algorithm on the GraphBig programs is quite good as
indicated in Table III and Table IV, considering that these are
large-scale application programs representative of data center
applications.

\section{Conclusion}
This paper describes an output-regulation technique inspired by Newton-Raphson's algorithm for solving algebraic
equations. The tracking controller has the form of an integrator with adjustable gain, designed for effective
regulation. The gain is adjusted in real time by simple computations in the feedback loop. Furthermore, the regulation
algorithm is robust to modeling uncertainties and computing errors in the loop, hence does not require
precise models of the plant.

We implemented the controller on Intel's  commodity microarchitecture, Haswell, in order to test it on various industry-benchmark programs. The control variable consists of the processor's clock rate, and the controlled quantity is the
spatial and temporal average of the cores' power. Due to the lack of adequate models for performance evaluation of
these systems, we programmed and performed a system-identification algorithm that is executed in real time.
We describe the main technical challenges  associated with  implementations of the controller.
Results of the experiments are presented and discussed in detail, and they exhibit fast
and effective convergence.
To the best of our knowledge, the paper presents the first implementation of a control law
at the core-level,  based on formal
control theory, and applied to  application programs that  are executed in large datacenters.


\begin{thebibliography}{99}

\bibitem{Wang11}
	X. Wang, K. Ma, and Y. Wang.
	Adaptive Power Control with Online Model Estimation for Chip Multiprocessors.
	\emph{IEEE Transactions on Parallel and Distributed Systems}, Vol. 22, pp. 1681-1696,  2011.




\bibitem{Bohrer02}
P. Bohrer, N.E. Elmootazbellah, T. Keller,  M. Kistler, C. Lefurgy,  C. McDowell, and R. Rajamony.
The casae for power management in web servers.
In {\it Power aware computing}, pp. 261-289, Springer, USA, 2002.

\bibitem{Hammarlund14}
P. Hammarlund, A. Martinez, A. Bajwa, D. Hill, E. Hallnor, H. Jiang, M. Dixon, M. Derr, M. Hunsaker, R. Kumar, R. Osborne, R. Rajwar, R.
Singhal, R. D'Sa, R. Chappell, S. Kaushik, S. Chennupaty, S. Jourdan, S. Gunther, T. Piazza, and T. Butron.  Haswell: The Fourth-Generation Intel Core Processor.  \emph{Micro}, Volume 34, pp. 6-20,  2014.

	
\bibitem{Shadron03}
	K. Skadron, M. R. Stan, W. Huang, S. Velusamy, K. Sankaranarayanan, and D. Tarjan.
	Temperature-aware microarchitecture.
	\emph{ACM SIGARCH Computer Architecture}, Vol. 31, pp. 2–13, 2003.
	
\bibitem{Liu12}
	G. Liu, M. Fan, and G. Quan.
	Neighbor-aware dynamic thermal management for multi-core platform.
	\emph{Design, Automation and Test in Europe Conference and Exhibition}, Dresden, Germany, March 12-16, 2012.


	
\bibitem{Yeo08}
	I. Yeo, C.C. Liu, and E.J. Kim.
	Predictive dynamic thermal management for multicore systems.
	\emph{Design Automation Conference}, Anaheim, CA, USA, June 8-13, 2008.
	
\bibitem{Kim15}
	S.W. Kim, T.M. Kim and C. Yoo.
	Workload prediction using run-length encoding for runtime processor power management.
	\emph{Electronics Letters}, Vol. 51, pp. 1759-1761,  2015.
	
	
\bibitem{Avirneni16}
	N. Avirneni, P. Ramesh, and A. Soman.
	Utilization Aware Power Management in Reliable and Aggressive Chip Multi Processors.
	\emph{IEEE TRANSACTIONS ON COMPUTERS}, Volume 65, pp. 979-991, 2016.
	
\bibitem{Raghavendra08}
	R. Raghavendra, P. Ranganathan, V. Talwar, Z. Wang, and X. Zhu.
	No power struggles: coordinated multi-level power management for the data center.
	\emph{SIGARCH Comput. Architecture News}, Volume 36, pp. 48-59, 2008.
	
\bibitem{Mishra10}
	A. Mishra, S. Srikantaiah, M. Kandemir, and C. Das.
	Cpm in cmps: Coordinated power management in chip-multiprocessors.
	\emph{Intl. Conference on High Performance Computing, Networking, Storage and Analysis}, New Orleans, LA, USA, Nov. 13-19, 2010.

\bibitem{Hellerstein04}
	J.L. Hellerstein, Y. Diao, S. Parekh, and D.M. Tilbury.  {\it Feedback Control of Computing Systems},
	John Wiley \& Sons,  2004.

	
\bibitem{Deval15}
	A. Deval, A. Ananthakrishnan, and C. Forbell.
	Power management on 14 nm Intel® Core− M processor.
	\emph{Low-Power and High-Speed Chips}, Yokohama, Japan, April 13-15, 2015.
	
\bibitem{Krishnaswamy15}
	V. Krishnaswamy, J. Brooks, G. Konstadinidis, C. McAllister, H. Pham, S. Turullols, J. Shin, Y. Gong, and H. Zhang.
	Fine-Grained Adaptive Power Management of the SPARC M7 Processor.
	\emph{Solid- State Circuits Conference}, San Francisco, CA,  Feb 22-26, 2015.
	
\bibitem{Chen13}
	H. Chen, A. Coskun, and M. Caramanis.
	Real-Time Power Control of Data Centers for Providing Regulation Service.
	\emph {IEEE Conference on Decision and Control}, Florence, Italy, Dec 10-13, 2013.

	
\bibitem{Lefurgy08}
	C. Lefurgy, X. Wang, and M. Ware.
	Power Capping: a Prelude to Power Shifting.
	\emph{Cluster Computing}, Volume 11, pp. 183 to 195, 2008.


\bibitem{Franklin14}
G.F. Franklin, J.D. Powell, and A. Emami-Naeini.
{\it Feedback Control of Dynamic Systems},
Prentice Hall, Pearson, 2014.

  \bibitem{Almoosa12}
N. Almoosa, W. Song, S. Yalamanchili, and Y. Wardi.
A Power Capping Controller for Multicore Processors.
\emph{Proc. American Control Conference}, Montreal, Canada, June 27-29, 2012.


	
\bibitem{Wardi16}
Y. Wardi, C. Seatzu, X. Chen, and S. Yalamanchili.
Performance Regulation of Stochastic Discrete Event Dynamic Systems Using Infinitesimal Perturbation Analysis.
  {\em Nonlinear Analysis: Hybrid Systems}, Volume 22, pp.116-136, November 2016.

    \bibitem{Almoosa12a}
N. Almoosa, W.   Song, Y.   Wardi,   and  S. Yalamanchili.
Throughput Regulation in Multicore Processors via IPA.
\emph{Proc. 51 IEEE Conference on Decision and Control (CDC),}
 Maui, Hawaii, December 10-13.

\bibitem{Chen15}
X. Chen, H. Xiao, Y. Wardi, and S. Yalamanchili.
Throughput Regulation in Shared Memory Multicore Processors.
{\em IEEE International Conference on High Performance Computing},  Bangalore, India, December 2015.

\bibitem{Seatzu14}
C. Seatzu and Y. Wardi.
 Performance Regulation Via Integral Control in a Class of Stochastic Discrete Event Dynamic Systems.
 {\it Proc. 12th International Workshop on Discrete Event Systems, WODES 2014}, Cachan, France, May 14-16, 2014.



\bibitem{Chen16}
 X. Chen, Y. Wardi, and S. Yalamanchili.
 IPA in the Loop: Control Design for Throughput Regulation in Computer Processors.
 {\em Proc. 13th Intl. Workshop on Discrete Event Systems (WODES'16)}, Xi'an, China, May 30 to June 1, 2016.

   \bibitem{Lancaster66}
P. Lancaster.
   Error analysis for the Newton-Raphson method.
     \emph{Numerische Mathematik},
     Vol. 9, pp. 55-68, 1966.

  \bibitem{Ho91}
Y.C. Ho and X.R. Cao.
{\it Perturbation Analysis of Discrete Event Dynamic Systems}, Kluwer Academic Publishers, Boston, Massachusetts, 1991.

\bibitem{Cassandras08}
C.G. Cassandras and S. Lafortune.
 {\it Introduction to Discrete Event Systems}, Springer Science+Business Media, New York, 2008.





\bibitem{Hennessy12}
J.L. Hennessy and D.A. Patterson.
 {\it Computer Architecture: A Quantitative
Approach}, Morgan Kaufmann, 2012.

\bibitem{Keesman11}
K. Keesman.
{\it  System identification: an introduction}.
 Springer Science \& Business Media, 2011.



\bibitem{Woo95}
S.C. Woo, M. Oharat, E.  Torriet, J. Singhi, and A. Guptat. The SPLASH-2 Programs:
Characterization and Methodological Considerations.
 {\it Proc.  22nd Annual
International symposium  on  Computer  architectures}, pp. 24-36,
S. Margherita Ligure, Italy, June 22-24, 2005.

\bibitem{Nai15}
L. Nai, Y. Xia, I. Tanase, H. Kimy, and C. Lin.
GraphBIG: Understanding Graph Computing in the Context of Industrial Solutions.
\emph{International Conference for High Performance Computing, Networking, Storage and Analysis},
Austin, TX,  Nov. 15-20, 2015.

\bibitem{Browne00}
S. Browne, J.  Dongarra, N. Garner, G. Ho, and P.  Mucci.
 A Portable Programming Interface for Performance Evaluation on Modern Processors.
   {\it The International Journal of High Performance Computing Applications}, Volume 14, number 3, pp. 189-204, Fall 2000.










\end{thebibliography}
\end{document}